\documentclass{ws-procs9x6}

\begin{document}

\title{CP violation in supersymmetric seesaw models}

\author{Junji Hisano}

\address{
ICRR, University of Tokyo\\
 5-1-5 Kashiwa-no-Ha\\
 Kashiwa, 277-8582, Japan }
\maketitle

\abstracts{In supersymmetric (SUSY) extensions of the seesaw mechanism
the neutrino Yukawa interaction induces the flavor and CP violating
sfermion mass terms via the radiative correction. In this article we
review the CP violating phenomena in the SUSY seesaw models.}  

\section{Introduction}

The seesaw mechanism\cite{seesaw}, which explains tiny neutrino
masses, is one of the most promising models beyond the standard model
(SM), after discovery of the neutrino oscillation. It is also
compatible with the matter unification in the GUTs. In the SO(10) GUT
quarks and leptons may be embedded into {\bf 16} dimensional
multiplets for each generation, in cooperation with the right-handed
neutrinos. In addition to it, the baryon number in the universe may be
explained by the leptogenesis\cite{Fukugita:1990gb}.  Hence, it is
important to search for signals in the models.

Supersymmetry (SUSY) is expected in the seesaw mechanism so that the
hierarchical structure is stabilized. If the SUSY breaking terms in
the minimal SUSY SM (MSSM) comes from dynamics whose energy scale is
above that of the seesaw mechanism or the extension, the neutrino Yukawa
interaction generates the flavor and CP violating SUSY breaking
terms\cite{Hall:1985dx}\cite{bm}. Thus, the flavor and CP violating
phenomena are sensitive to the SUSY seesaw models.

This paper is concentrated into the CP violating aspects in the SUSY
seesaw model and the extension to the SUSY GUT. The lepton flavor
violating (LFV) processes and the related topics in the models are
reviewed by Masiero in this volume.

\section{Minimal SUSY seesaw model}

In the minimal SUSY seesaw model, in which only three right-handed
neutrinos are introduced, the SUSY breaking terms in the lepton
sector are sensitive to the neutrino Yukawa coupling. Thus, the CP
violation in the minimal SUSY seesaw model may lead to the T-odd
asymmetry in the lepton-flavor violating lepton decay, such as
$\mu\rightarrow 3e$, and the leptonic EDMs. In this section, we review
the flavor structure in the leptonic SUSY breaking terms, which is
predicted in the model, and the CP violating observables.

In the minimal SUSY seesaw model the relevant leptonic part of the
superpotential is
\begin{eqnarray}
W_{\rm seesaw} & =& f^\nu_{ij} L_i \overline{N}_j  \overline{H}_f
  +  f^l_{ij} \overline{E}_i  L_j H_f 
+ \frac{1}{2}{M}_{ij} {\overline{N}}_i \overline{N}_j 
\label{MseesawM}
\end{eqnarray}
where the indexes $i,j$ run over three generations and ${M}_{ij}$
is the heavy singlet neutrino mass matrix.  In addition to the three
charged lepton masses, this superpotential has eighteen physical
parameters, including six real mixing angles and six CP-violating
phases, because the Yukawa coupling and the Majorana mass matrices
are given after removing unphysical phases as 
$f^l_{ij} = f_{l_i} \delta_{ij}$,
$f^\nu_{ij} = X^\star_{ik} f_{\nu_k} 
{\rm e}^{-i \varphi_{\nu_k}} W^\star_{kj} 
{\rm e}^{-i \overline{\varphi}_{\nu_k}}$, and 
$M_{ij} =\delta_{ij} M_{N_k}.$
Here, $\sum_i \varphi_{\nu_i}=0$ and  
$\sum_i \overline{\varphi}_{\nu_i}= 0$, and each $W$ and $X$ are 
unitary matrices with one phase. 

At low energies the effective theory after integrating out the right-handed
neutrinos is given by the effective superpotential,
\begin{eqnarray}
W_{\rm eff} &=&  f_{l_{i}} \overline{E}_i  L_i H_f 
  + \frac{1}{2 v^2 \sin^2\beta} ({m_\nu})_{ij} (L_i \overline{H}_f)(L_j \overline{H}_f) \,,
\label{weff}
\end{eqnarray}
where we work in a basis in which the charged lepton Yukawa couplings
are diagonal. The second term in (\ref{weff}) leads to the light
neutrino masses and mixings. The explicit form of the small neutrino
mass matrix $({m_\nu})$ is given by
$({m_\nu})_{ij} =
\sum_k {f^\nu_{ik}f^\nu_{jk}} v^2 \sin^2\beta/
            {{M}_{N_k}}$.
The light neutrino mass matrix $({m_\nu})$ is
symmetric, with nine parameters, including three real mixing angles
and three CP-violating phases. It can be diagonalized by a unitary matrix
$Z$ as $Z^T {m}_\nu Z = {m}^D_\nu$.
By redefinition of fields one can rewrite $Z \equiv U P,$ where $P
\equiv {\rm diag}(e^{i\phi_1}, e^{i\phi_2}, 1 )$ and $U$ is the MNS
matrix, with the three real mixing angles and the remaining
CP-violating phase.

The abilities to probe the minimal seesaw model by the low-energy
neutrino experiments, such as the neutrino oscillation and the double
$\beta$ decay, are limited. Nine parameters associated with the
heavy-neutrino sector cannot be measured in a direct way.  However, we
may probe the model by phenomena induced by the SUSY breaking terms in
the MSSM.

If the SUSY-breaking terms are generated above the right-handed
neutrino mass scale, the renormalization effects may induce sizable
LFV slepton mass terms, which lead to the LFV charged lepton decays.
If the SUSY-breaking parameters at the GUT scale or the Planck scale
are universal, off-diagonal components in the left-handed slepton mass
matrix $(m^2_{\tilde{l}_L})$ and the trilinear slepton coupling $(A_l)$ take the
approximate forms, 
\begin{eqnarray}
(m_{{\tilde{l}_L}}^2)_{ij}&\simeq&
-\sum_k \frac{f_{ik}^{\nu\star}f_{jk}^{\nu}}{16\pi^2}
\left[m_0^2(3\log\frac{M_G^2}{M_{N_k}^2}+1)
      +A_0^2(\log\frac{M_G^2}{M_{N_k}^2}+1)
\right]
\,,
\nonumber\\
( A_l)_{ij} &\simeq&
-\sum_k f_{l_i} A_0  \frac{f_{ik}^{\nu\star}f_{jk}^{\nu}}{16\pi^2} 
\left[\log\frac{M_G^2}{M_{N_k}^2}+1
\right]
 \,,
\label{leading}
\end{eqnarray}
where $ i\ne j$, and the off-diagonal components of the right-handed
slepton mass matrix are suppressed. Here, we include the one-loop
finite parts\cite{Farzan:2004qu} in the radiative correction, and
ignore terms of higher order in $f_l$, assuming that $\tan\beta$ is
not extremely large.

The non-vanishing off-diagonal components in $(m_{\tilde{l}_L}^2)$ and $(A_l)$
predict the charged lepton-flavor violating decays, whose measurements
supply the information about the seesaw model, which is independent of
the low-energy neutrino experiments. Ignoring the one-loop finite
corrections, the terms in Eq.~(\ref{leading}) are proportional to
$H_{ij}
=
\sum_k 
{f^{\nu\star}_{ik}}
{f^{\nu}_{jk}}
\log({M^2_{G}}/{{M}^2_{N_k}})$.
Here, the Hermitian matrix $H$, whose diagonal terms are real and
positive, is defined in terms of $f^\nu$ and the heavy neutrino masses
${M}_{N_k}$.  This matrix has nine parameters including three phases,
which are clearly independent of the parameters in $({m}_\nu)$. Thus
two matrices $({m}_\nu)$ and $H$ together provide the required
eighteen parameters, including six CP-violating phases, by which we
can parameterize the minimal SUSY seesaw model\cite{Ellis:2002fe}. The
off-diagonal terms, $H_{ij}(i\ne j)$, are related to the LFV $l_i-l_j$
transition, and they are related to the LFV charged lepton decays.

Now we discuss the CP violation in the minimal SUSY seesaw model.
The CP violating observables of the charged leptons are the T-odd
asymmetry in $l\rightarrow 3l'$ and the leptonic EDMs. 

First, we discuss the T-odd asymmetry in $\mu^+\rightarrow
e^+e^-e^+$. In order for CP violation to appear in any process,
interference between different terms in the amplitude for the process
must occur.  Therefore, all possible observables in $l\to l'\gamma$
decays, such as differences between the $\mu^{+}
\rightarrow e^{+}\gamma$ and $\mu^{-} \rightarrow e^{-}\gamma$ rates,
vanish in the leading order of perturbation theory. Moreover, the
process $\mu^{-} \rightarrow e^{-}\gamma$ is not measurable with high
accuracy because of the large backgrounds. However, when muons are
polarized, a T-odd asymmetry for final-state particles in $\mu^{+}
\rightarrow e^{+}e^{+}e^{-}$ can be defined. Since CPT is conserved, the
T-odd asymmetry measures the amount of CP violation in the model.

The muon polarization vector $\vec{P}$ can be defined in the
coordinate system in which the $z$ axis is taken to be the direction
of the electron momentum, the $x$ axis the direction of the most
energetic positron momentum, and the $(z\times x)$ plane defines the
decay plane perpendicular to the $y$ axis.  The T-odd asymmetry is
then defined\cite{Okada} by
\begin{eqnarray}
A_{T} &=& \frac{N(P_{\it y} >0)-N(P_{\it y} <0)}
{N(P_{\it y} >0)+N(P_{\it y} <0)},
\label{AT}
\end{eqnarray}
where $N(P_i >(<)0)$ 
denotes the number of events with a positive (negative) $P_i$ 
component for the muon polarization. $A_T$ is limited to be below
$24\%$\cite{deGouvea:2000cf}.

In the branching ratio for $\mu
\rightarrow 3e$, the contribution from photonic penguin diagram tends 
to dominate due to the phase-space integral while the $Z$ penguin and
box diagrams also gives the contribution. Then, assuming that the
photonic penguin diagram dominates in ${Br}(\mu^{+} \rightarrow
e^{+}e^{+}e^{-})$, the T-odd asymmetry $A_T$, induced by non-vanishing
off-diagonal terms in $(m_{L}^2)$, is approximately
given\cite{Ellis:2001xt} by
\begin{eqnarray}
A_T&=&\frac{{\rm Im}\Big[(\Delta_{12}^{{l}})_L (\Delta_{23}^{{l}})_L
(\Delta_{31}^{{l}})_L \Big]}{|(\Delta_{12}^{{l}})_L|^2}
\nonumber\\
&&\times
\frac{0.039+0.196 \tan\beta + 0.017/\tan\beta }
     {
\left|(1+2.4\tan\beta)
-
\frac{(\Delta_{23}^{{l}})_L (\Delta_{31}^{{l}})_L}{(\Delta_{21}^{{l}})_L}
(0.64 +1.12\tan\beta)
\right|^2
}\,,
\label{approxAT}
\end{eqnarray}
where 
$(\Delta^{{l}}_{ij})_L \equiv ({(m_{
L}^2)_{ij}}/{m_S^2})$.
Here, in writing Eq.~(\ref{approxAT}), we have taken $(A_l)_{ij}=0$
($i\ne j$), for simplicity.  We see explicitly how $A_T$ in
Eq.~(\ref{approxAT}) depends on the Jarlskog invariant $J_{
L}=\mbox{\rm{Im}} [ (m_{\tilde{l}_L}^2)_{12} (m_{\tilde{l}_L}^2)_{23}
(m_{\tilde{l}_L}^2)_{31} ]$. It is found that $A_T$ could in principle
reach $\sim 10\%$.  However, if ${\rm Im}[(\Delta_{12}^{{l}})_L
(\Delta_{23}^{{l}})_L (\Delta_{31}^{{l}})_L] \ll
|(\Delta_{12}^{{l}})_L|^2$, as one might expect, or if $\tan\beta\gg
1$, $A_T$ is suppressed.

\begin{figure}[ht]
\centerline{
\epsfxsize=2.0in\epsfbox{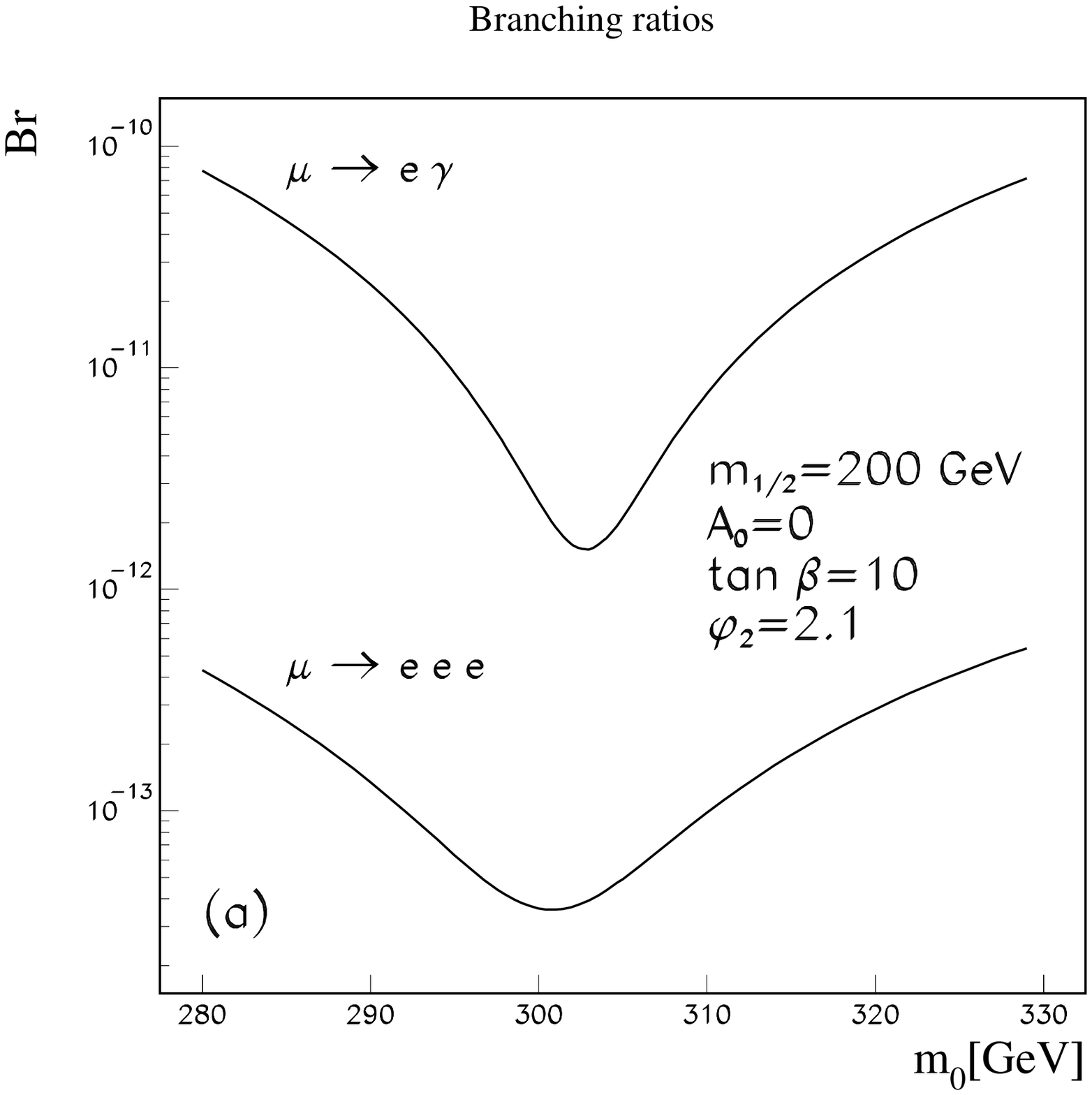} 
\epsfxsize=2.0in\epsfbox{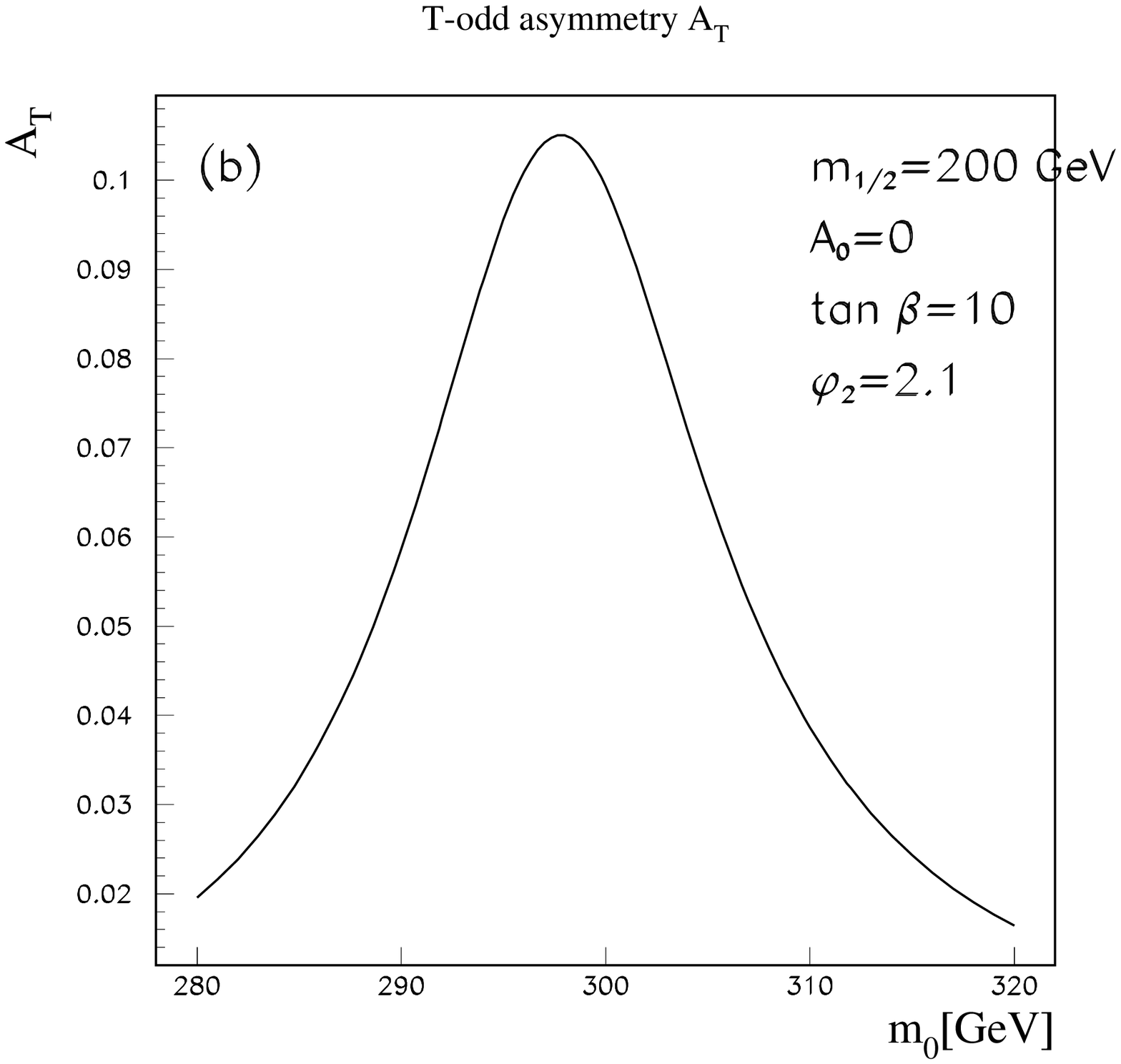} 
}   
\caption{(a) Branching ratios for the decays
$\mu \rightarrow e\gamma$ and $\mu \rightarrow 3 e$
and (b) the T-odd asymmetry $A_T$ in $\mu^{+} \rightarrow e^{+}e^{+}e^{-}$
decay, as functions of the common soft mass $m_0$.}
\label{fig1}
\end{figure}

In Fig.~\ref{fig1} we show (a) the branching ratios for the decays
$\mu
\rightarrow e\gamma$ and $\mu \rightarrow 3 e$ and
(b) the T-odd asymmetry $A_T$ in $\mu^{+} \rightarrow e^{+}e^{+}e^{-}$
decay, as functions of the common soft mass $m_0$. For the neutrino
parameters and others in the minimal SUSY seesaw model, see
Ref.~\cite{Ellis:2001xt}. When the branching ratio for $\mu
\rightarrow e\gamma$ is suppressed due to the accidental cancellation,
the T-odd asymmetry is enhanced. This is because the penguin 
contribution is suppressed and and becomes comparable to the other 
contributions. 

The leptonic EDMs depend on the flavor-diagonal Jarlskog invariants.
If the source of the flavor violation comes from only the Yukawa
interactions in the minimal SUSY seesaw model, the flavor-diagonal
Jarlskog invariants, which consist of the Yukawa coupling constants,
are given as
\begin{eqnarray}
J^{(i)}_{\rm edm} &=& \mbox{\rm{Im}} \left[  f^{l\star}\left[f^{\nu}
f^{\nu\dagger}
\left[f^{lT} f^{l\star},~ f^\nu f^{\nu \dagger} \right] 
f^{\nu}f^{\nu\dagger} \right]_{ii}\right].
\end{eqnarray}
These are much suppressed by the Yukawa coupling constants. This
situation is similar to the hadronic EDMs in the SM, where the CP
violation comes from the CKM matrix. However, in the minimal SUSY
seesaw model, when the right-handed neutrinos are not degenerate in
mass, we have other flavor-diagonal Jarlskog invariants,
\begin{eqnarray}
J^{(i)\prime }_{\rm edm} &=& \mbox{\rm{Im}} \left[  f^{l\star}\left[f^{\nu}
f^{\nu\dagger}, f^{\nu} \log\frac{M_G^2}{M_{N}^2} f^{\nu\dagger}\right]_{ii}
\right] .
\end{eqnarray}
Obviously, they vanish when the right-handed neutrino masses are
degenerate.  Thus, it is expected that the leptonic EDMs are enhanced
significantly when the right-handed neutrinos masses are not
degenerate\cite{Ellis:2001yz}.

When the SUSY breaking terms are generated as in Eq.~(\ref{leading}),
the non-trivial Jarlskog invariants, $J^{(i)\prime }_{\rm edm}$, are
given as\cite{Farzan:2004qu}  
\begin{eqnarray}
\mbox{\rm{Im}} \left[\left[( A_l)( m_{\tilde{l}_L}^2)\right]_{ii}
\right]&=& \frac{A_0m_0^2}{(4\pi)^4}J^{(i)\prime }_{\rm edm},
\end{eqnarray}
and the leptonic EDMs are proportional to them. It is shown that the
electron and muon EDMs may reach to $10^{-29}$ and $10^{-27} e cm$,
respectively, for $m_S\sim 200$GeV\cite{Farzan:2004qu}.  The CP
violating phases contributing to the leptonic EDMs are independent of
those in other CP violating observables, such as the T-odd asymmetry
in $\mu \rightarrow 3 e$ or the neutrino oscillation.

\section{SUSY SU(5) GUT with right-handed neutrinos}

When the SUSY seesaw model is extended to the SUSY GUTs, the neutrino
Yukawa interaction also induces the flavor violation in the hadronic
sector via the radiative corrections to the squark SUSY breaking
terms. In this section, we discuss the hadronic CP violation in the
SUSY GUTs with the right-handed neutrinos. Here, we take the SUSY
SU(5) GUT for simplicity. In this model, doublet leptons and
right-handed down-type quarks are embedded in common {\bf
5}-dimensional multiplets, while doublet quarks, right-handed up-type
quarks, and right-handed charged leptons are in the {\bf
10}-dimensional ones. Thus, the neutrino Yukawa coupling induces the
right-handed down-type squark mixing, and this leads to rich flavor
and CP violating phenomena in hadrons. They are also correlated due to
the GUT relation, especially processes of the second and third
generation transition\cite{Hisano:2003bd}.

First, we review the flavor structure in the squark and slepton mass
matrices in the SUSY SU(5) GUT with the right-handed neutrinos. The
Yukawa interactions for quarks and leptons and the Majorana mass terms
for the right-handed neutrinos in this model are given by the following
superpotential,
\begin{equation}
W= 
\frac14 f_{ij}^{u} \Psi_i \Psi_j H 
+\sqrt{2} f_{ij}^{d} \Psi_i \Phi_j \overline{H}
+f_{ij}^{\nu} \Phi_i \overline{N}_j {H}
+M_{ij} \overline{N}_i \overline{N}_j,
\label{superp_gut}
\end{equation}
where $\Psi$ and $\Phi$ are for {\bf 10}- and {$\bf \bar{5}$}-dimensional
multiplets, respectively, and $\overline{N}$ is for the right-handed
neutrinos.  $H$ ($\overline{H}$) is {\bf 5}- ({$\bf \bar{5}$}-)
dimensional Higgs multiplets.  After removing the unphysical degrees
of freedom, the Yukawa coupling constants in Eq.~(\ref{superp_gut})
are given as 
$f^u_{ij} = 
V_{ki} f_{u_k} {\rm e}^{i \varphi_{u_k}}V_{kj}$,
$f^d_{ij} =f_{d_i} \delta_{ij}$, and 
$f^\nu_{ij} = {\rm e}^{i \varphi_{d_i}} 
U^\star_{ij} f_{\nu_j}$.
Here, $\varphi_{u}$ and $\varphi_{d}$ are CP-violating phases inherent
in the SUSY SU(5) GUT. They satisfy $\sum_i \varphi_{f_i} =0$
$(f=u$ and $d)$.  The unitary matrix $V$ is the CKM matrix in the extension
of the SM to the SUSY SU(5) GUT, and each unitary matrices $U$ and $V$
have only a phase. When the Majorana mass matrix for the right-handed
neutrinos is diagonal, $U$ is the MNS matrix
observed in the neutrino oscillation.  In this section we  assume the
diagonal Majorana mass matrix in order to avoid the complexity due to
the structure. 

The colored Higgs multiplets $H_c$ and $\overline{H}_c$ are introduced
in $H$ and $\overline{H}$ as SU(5) partners of the Higgs doublets in
the MSSM, respectively. They have new flavor-violating interactions in
Eq.~(\ref{superp_gut}). If the SUSY-breaking terms in the MSSM are
generated by dynamics above the colored Higgs masses, such as in the
gravity mediation, the sfermion mass terms may get sizable corrections
by the colored Higgs interactions.  In the minimal supergravity
scenario the SUSY breaking terms are supposed to be given at the
reduced Planck mass scale ($M_G$). In this case, the flavor-violating
SUSY breaking mass terms at low energy are induced by the radiative
correction, and they are qualitatively given in a flavor basis as
\begin{eqnarray}
(m_{{\tilde{u}_L}}^2)_{ij}  &\simeq&-
V_{i3}V_{j3}^\star  \frac{f_{b}^2}{(4\pi)^2}
\;\; (3m_0^2+ A_0^2) \;\; 
(2 \log\frac{M_G^2}{M_{H_c}^2}+ \log\frac{M_{H_c}^2}{m^2_{S}}),\nonumber\\
(m_{\tilde{u}_R}^2)_{ij}  &\simeq& -
{\rm e}^{-i\varphi_{u_{ij}}} 
V_{i3}^\star V_{j3} \frac{2f_{b}^2}{(4\pi)^2}
\;\; (3m_0^2+ A_0^2) \;\; 
\log\frac{M_G^2}{M_{H_c}^2}, \nonumber\\
(m_{{\tilde{d}_L}}^2)_{ij}  &\simeq&-
V_{3i}^\star
V_{3j} \frac{f_{t}^2}{(4\pi)^2} 
\;\; (3m_0^2+ A_0^2) \;\; 
(3 \log\frac{M_G^2}{M_{H_c}^2}+ \log\frac{M_{H_c}^2}{m_{S}^2}),\nonumber\\
(m_{\tilde{d}_R}^2)_{ij}  &\simeq&-
{\rm e}^{i\varphi_{d_{ij}}}  U^\star_{ik}U_{jk} 
\frac{f_{\nu_k}^2}{(4\pi)^2} 
\;\; (3m_0^2+A_0^2) \;\; 
\log\frac{M_G^2}{M_{H_c}^2},\nonumber\\
(m_{\tilde{l}_L}^2)_{ij}  &\simeq&-
U_{ik}U_{jk}^\star
\frac{f^2_{\nu_k} }{(4\pi)^2} 
\;\; (3m_0^2+ A_0^2) \;\; 
\log\frac{M_G^2}{M_{N_k}^2},\nonumber\\
(m_{{\tilde{e}_R}}^2)_{ij}  &\simeq&-
{\rm e}^{i\varphi_{d_{ij}}} 
V_{3i}V^\star_{3j}
\frac{3 f_{t}^2}{(4\pi)^2} 
\;\; (3m_0^2+ A_0^2)\;\; 
\log\frac{M_G^2}{M_{H_c}^2},
\label{sfermionmass}
\end{eqnarray}
with $i\ne j$, where
$\varphi_{u_{ij}}\equiv\varphi_{u_{i}}-\varphi_{u_{j}}$ and
$\varphi_{d_{ij}}\equiv\varphi_{d_{i}}-\varphi_{d_{i}}$ and $M_{H_c}$
is the colored Higgs mass.  $f_t$ is the top quark Yukawa coupling
constant while $f_b$ is for the bottom quark. As mentioned above, the
off-diagonal components in the right-handed squarks and slepton mass
matrices are induced by the colored Higgs interactions, and they
depend on the CP-violating phases in the SUSY SU(5) GUT with the
right-handed neutrinos\cite{Moroi:2000tk}.

When both the left-handed and right-handed squarks have the
off-diagonal components in the mass matrices, the EDMs and CEDMs for
the light quarks are enhanced significantly by the heavier quark
mass\cite{Dimopoulos:1994gj}\cite{Barbieri:1995tw}.  The EDMs and
CEDMs contribute to the hadronic EDMs, and they are constrained by the
observations.  In the SUSY SU(5) GUT with the right-handed neutrinos,
the neutrino Yukawa coupling induces the flavor-violating mass terms
for the right-handed down-type squarks.  The flavor-violating mass
terms for the left-handed down-type squarks are expected to be
dominated by the radiative correction, which is controlled by the CKM
matrix, induced by the top quark Yukawa coupling as in
Eq.~(\ref{sfermionmass}). Then, we can investigate or constrain the
structure in the neutrino sector by the hadronic EDMs, which is
generated by the CEDMs and EDMs of the down and strange
quarks.\footnote{ The up quark EDM and CEDM are also predicted in this
model.  Since they are suppressed by the bottom quark Yukawa coupling,
they may be observable in near future when $\tan\beta$ is
large\cite{Romanino:1996cn}.}

The CEDMs for the light quarks, including the strange quark,
contribute to the hadronic EDMs since the CP-violating nucleon
coupling is induced by them. Here, we consider only the CEDMs for the
light quarks.  While the EDMs for the up and down quarks contribute to
the neutron EDM, it is found that the EDM contributions are comparable to
or smaller than the CEDM contributions.

The CEDMs of the down-type light quarks derived by the flavor
violation in the both the left-handed and right-handed quark mass
matrices are given by the following dominant contribution, which is
enhanced by the heavier quark masses,
\begin{eqnarray}
 {d}^C_{d_i} &=& \frac{\alpha_s}{4\pi}\frac{m_{\tilde g}}{\overline{m}^2_{\tilde d}}
f\left(\frac{m_{\tilde{g}}^2}{\overline{m}^2_{\tilde{d}}}
\right)  \sum_j {\mbox{\rm{Im}}}
\left[(\Delta^d_{ij})_{L}(\Delta^d_{j})_{LR}(\Delta^d_{ji})_{R}
\right],
\label{SUSYEDM}
\end{eqnarray}
where $m_{\tilde{g}}$ and $\overline{m}_{\tilde{d}}$ are the gluino
and averaged squark masses.
The mass insertion parameters are defined as
$(\Delta^d_{ij})_{L/R}\equiv
(m_{{\tilde{d}_{L/R}}}^2)_{ij}/{\overline{m}_{{\tilde{f}}}^2}$ and
$( \Delta_{i}^{d})_{LR} 
\equiv  {m_{d_i}(A_i^{(d)} -\mu\tan\beta)}/{\overline{m}^2_{\tilde{d}}}$.
The function $f(x)$ is given in Ref.~\cite{Hisano:2004pw} 
and $f(1) =-11/180$. 

In order to translate the CEDMs of the light quarks to the EDMs of 
neutron and ${^{199}}$Hg atom,  we use the evaluation of 
the  EDMs of neutron
and ${^{199}}$Hg atom in Ref.~\cite{HS2},
\begin{eqnarray}
d_n &=& -1.6 \times e (d_u^C+0.81\times d_d^C+0.16\times d_s^C),
\nonumber\\
d_{\rm Hg}&=&-8.7\times 10^{-3}\times e(d_u^C-d_d^C+0.005 d_s^C),
\end{eqnarray}
where $d_n$ is generated by the charged meson loops and $d_{\rm Hg}$
comes from the nuclear force by the pion exchange
in the chiral perturbation theory.  The
experimental upperbounds on the EDMs of neutron\cite{Harris:jx} and
$^{199}$Hg atom\cite{Romalis:2000mg} are
 $|d_{n}| < 6.3 \times 10^{-26} e\, cm$ and 
 $|d_{\rm Hg}| < 1.9\times 10^{-28} e\, cm$, 
respectively (90\%C.L.). Thus, the upperbounds on the quark CEDMs are
$e|{d}^C_u|<3.9(2.2)\times 10^{-26}\;e\;cm$, 
$e|{d}^C_d|<4.8(2.2)\times 10^{-26}\;e\;cm$, and
$e|{d}^C_s|<2.4(44) \times 10^{-25}\;e\;cm$,
from the EDM of neutron ($^{199}$Hg atom). Here, we assume that the
accidental cancellation among the CEDMs does not suppress the EDMs.
The constraint on ${d}^C_s$ from $^{199}$Hg atom is one-order weaker
than that from neutron, since the contribution to the EDM of $^{199}$Hg
atom is suppressed by $\pi^0$-$\eta^0$ mixing\cite{HS2}. 

\begin{figure}[ht]
\centerline{
\epsfxsize=2.0in\epsfbox{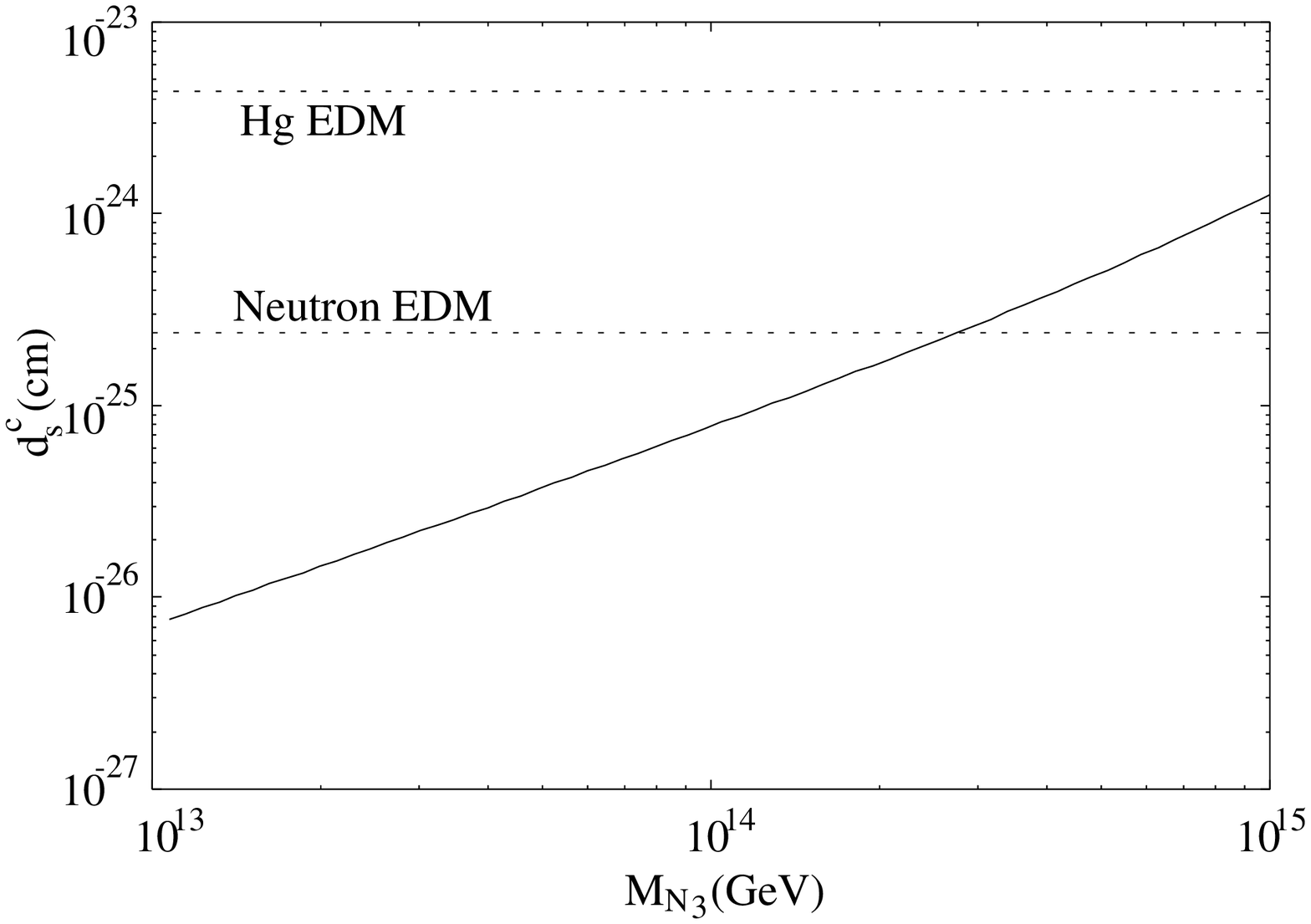} 
\epsfxsize=2.0in\epsfbox{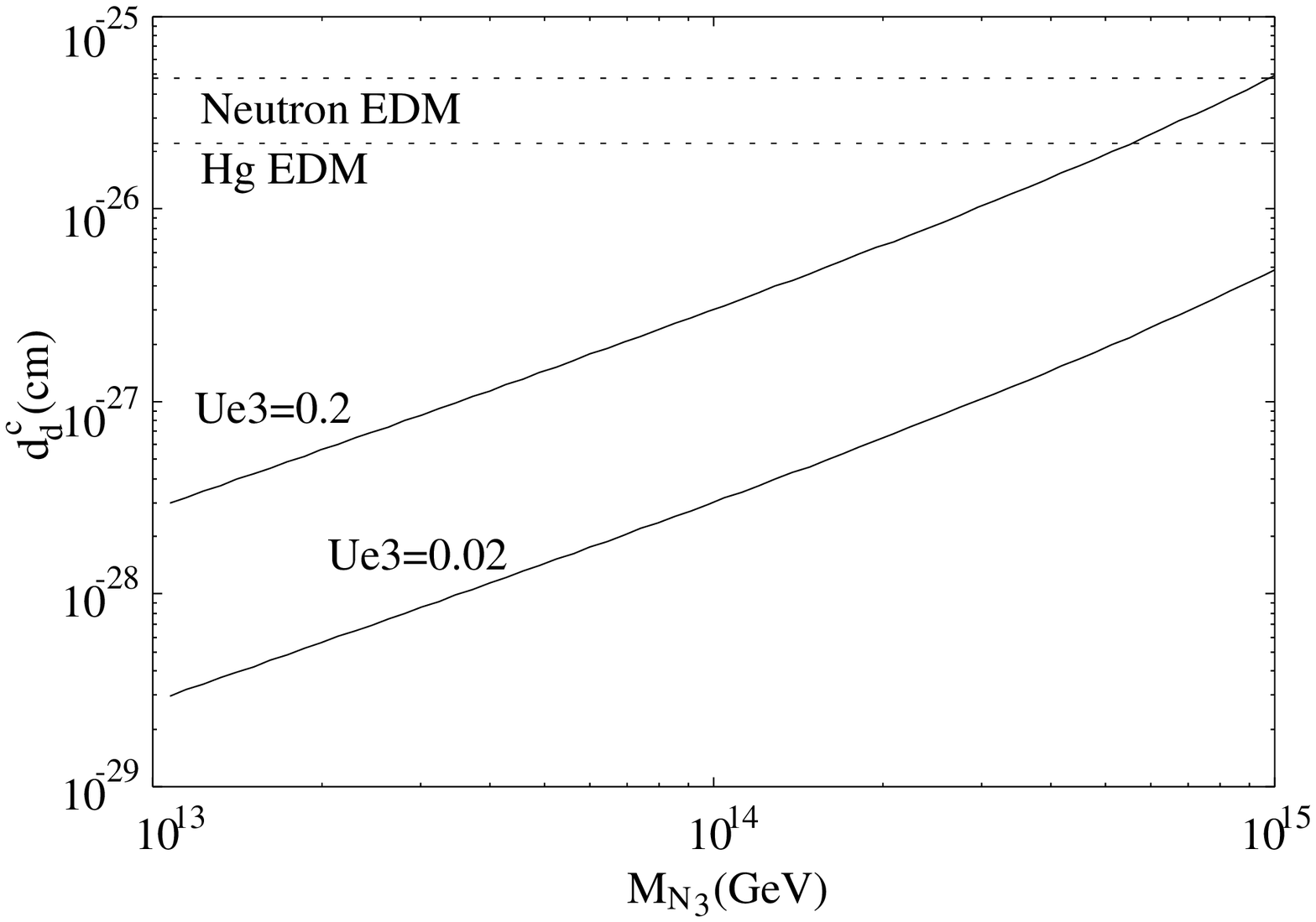}}   
\caption{CEDMs for the strange quark in (a) and for the down quark in (b) as
functions of the right-handed tau neutrino mass, $M_{N_3}$.  Here,
$M_{H_c}=2\times 10^{16}$GeV, $m_{\nu_\tau}=0.05$eV,
$U_{\mu3}=1/\sqrt{2}$, and $U_{e3}=0.2$ and 0.02.  For the MSSM
parameters, we take $m_0=500$GeV, $A_0=0$, $m_{\tilde{g}}=500$GeV and
$\tan\beta=10$.}
\label{fig2}
\end{figure}

In Fig.~\ref{fig2} the CEDMs for the down and strange quarks in the
SUSY SU(5) GUT with the right-handed neutrinos are
shown\cite{Hisano:2004pw}. The minimal supergravity scenario is
assumed and $M_{H_c}=2\times 10^{16}$GeV. In Fig.~\ref{fig2}(a) the
strange quark CEDM is presented as a function of the right-handed tau
neutrino mass. We take $m_{\nu_\tau}=0.05$eV and
$U_{\mu3}=1/\sqrt{2}$.  For the SUSY breaking parameters,
$m_0=500$GeV, $A_0=0$, $m_{\tilde{g}}=500$GeV and $\tan\beta=10$,
which lead to $\overline{m}_{\tilde{q}}\simeq 640$GeV. The
contributions from the electron and muon neutrino Yukawa interactions
to the flavor violation in the right-handed down-type squark mass
matrix are ignored. The contributions are bounded by the constraints
from the $K^0$--$\overline{K}^0$ mixing and $Br(\mu\rightarrow
e\gamma)$ when $|U_{e2}|\sim 1/\sqrt{2}$.  From this figure, the
right-handed tau neutrino mass should be smaller than $\sim 3\times
10^{14}$GeV.  In Fig.~\ref{fig2}(b), the down quark CEDM is presented
as a function of the right-handed tau neutrino mass. This comes from
non-vanishing $U_{e3}$ in our assumption that the right-handed
neutrino mass matrix is diagonal. The current bound is not significant
even when $U_{e3}=0.2$.

The new technique for the measurement of the deuteron EDM has a great
impact on the quark CEDMs if it is
realized\cite{Semertzidis:2003iq}. If they establish the sensitivity
of $d_D
\sim 10^{-27}e\,cm$, we may probe the new physics to the level of
$e d^C_s \sim 10^{-26}\;e\;cm$ and $e d^C_d \sim e d^C_u \sim
10^{-28}\;e\;cm$, which are much stronger than the bounds from the
neutron and $^{199}$Hg atom EDMs\cite{HS2}. This may imply that we may
probe the structure in the neutrino sector even if $M_{N_3}\sim
10^{13}$GeV or $U_{e3} \sim 0.02$.

Finally, we discuss the correlation between the hadronic EDM and the
CP asymmetry in $B\rightarrow \phi K_s$ ($S_{\phi K_s}$). The
$b\rightarrow s \bar{s} s$ transition is sensitive to the new physics
since it is induced at one-loop level\cite{gw}. When the right-handed
down-type squarks have the flavor mixing, the $b$--$s$ penguin diagram
may give a sizable contribution in $B\rightarrow \phi K_s$. Thus,
if the deviation from the SM prediction in $B\rightarrow \phi K_s$ is
observed, it might be a signature of the effect of the neutrino Yukawa
coupling in the SUSY GUTs with the right-handed neutrinos.  On the
other hand, the strange quark CEDM is also generated in the model, and
they have a strong correlation\cite{HS}.

The $b$--$s$ penguin contribution induced by the right-handed squark
mixing is dominated by the gluino diagram, and it is represented by
the effective operator $ H= - C_8^{R} ({g_s}/{(8\pi^2)})
m_b\overline{s_R}(G\sigma) b_L$.  When both the left-handed and
right-handed down-type squarks have flavor violation, we get a strong
correlation between $C_8^R$ and $d_s^C$ as
\begin{eqnarray}
{d}_s^C &=& -\frac{m_b}{4\pi^2} \frac{11}{21}
{\mbox{\rm Im}}\left[( \delta_{LL}^{(d)})_{23} C_8^{R}\right]
\label{massin}
\end{eqnarray}
up to the QCD correction. Here, we take
$m_{\tilde{g}}=\overline{m}_{\tilde{d}}$. The coefficient $11/21$ in
Eq.~(\ref{massin}) changes from 1 to $1/3$, depending on the SUSY mass
spectrum.

\begin{figure}[ht]
\centerline{
\epsfxsize=2.0in\epsfbox{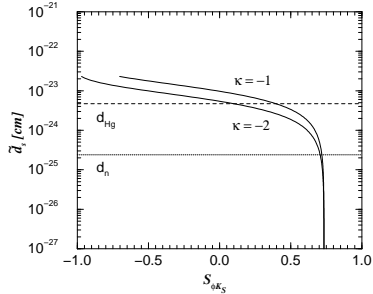} 
}   
\caption{
The correlation between ${d}^C_s$ and $S_{\phi K_s}$. $\kappa$ comes from 
the matrix element of chromomagnetic
moment  in $B\rightarrow \phi K_s$.
The dashed (dotted) line is the upperbound on ${d}^C_s$ from the EDM of 
$^{199}$Hg atom (neutron). 
}
\label{fig3}
\end{figure}

In Fig.~\ref{fig3}, the correlation between ${d}^C_s$ and $S_{\phi
K_s}$ is shown, assuming Eq.~(\ref{massin}). The detail of this figure is
shown in Ref.~\cite{HS2}. Here, we take $(\Delta_{23}^{d})_{L}
=-0.04$, ${\mathrm{arg}}[C_8^{R}]=\pi/2$ and $|C_8^R|$ corresponding
to $10^{-5}<|(\Delta_{32}^{d})_{R}|<0.5$. $\kappa$ is a parameter for
the matrix element of chromomagnetic moment in $B\rightarrow \phi
K_s$, and we show the results for $\kappa=-1$ and $-2$.  From this
figure, the deviation of $S_{\phi K_s}$ from the SM prediction due to
the gluon penguin contribution should be suppressed when the
constraints on ${d}^C_s$ from the $^{199}$Hg atomic and the neutron
EDMs are applied.  We find that the neutron EDM gives a stronger bound
on $S_{\phi K_s}$. Moreover, $S_{\phi K_s}$ may be constrained further
by the future deuteron EDM measurements.  Therefore, the hadronic EDMs
give a very important implication to $S_{\phi K_s}$.  Of cause, it
should be careful to compare the EDM constraints with other low-energy
observables, which are theoretically controlled better, since the
hadronic EDMs may suffer from more hadronic uncertainties. However,
the orders of the magnitude in the EDM constraints are still expected
to have the significance.

\section{Summary}

In this article we review the CP violating phenomena in the SUSY
seesaw models. In the SUSY extensions of the seesaw mechanism the
neutrino Yukawa interaction induces the flavor and CP violating
sfermion mass terms via the radiative correction. Thus, we may probe
the models by studying the flavor and CP violating phenomena.

The recent results for the $b$-$s$ penguin processes, including
$B\rightarrow \phi K_s$, in the BABAR and BELLE experiments are
converging. The combined results of the various $b$-$s$ penguin
processes are $2.7\sigma$ and $2.4\sigma$ deviated from the SM
prediction in the BABAR\cite{babar} and BELLE\cite{belle} experiments,
respectively. Though it is premature to judge whether they are signals
of new physics, it is important to discuss the sensitivity to new
physics. The deviation may be explained by introduction of the
right-handed bottom and strange squarks mixing, such as in the SUSY
SU(5) GUT with the right-handed neutrinos, however, the hadronic EDM
constraints give bounds on the deviation. Even if the discrepancy
comes from the hadronic uncertainties, the further improvements of the
bound on the hadronic EDMs and the $b$-$s$ penguin processes are very
important.

%
%
%
%


\begin{thebibliography}{0}
\bibitem{seesaw} M. Gell-Mann, P. Ramond and R. Slansky, Proceedings of   
the Supergravity Stony Brook Workshop, New York, 1979, eds. P. Van
Nieuwenhuizen and D. Freedman (North-Holland, Amsterdam); T. Yanagida,
Proceedings of the Workshop on Unified Theories and Baryon Number in
the Universe, Tsukuba, Japan 1979 (edited by A.  Sawada and A.
Sugamoto, KEK Report No.  79-18, Tsukuba).

\bibitem{Fukugita:1990gb}
M.~Fukugita and T.~Yanagida,
Phys.\ Rev.\ D {\bf 42} (1990) 1285.

\bibitem{Hall:1985dx}
L.~J.~Hall, V.~A.~Kostelecky and S.~Raby,
Nucl.\ Phys.\ B {\bf 267} (1986) 415.

\bibitem{bm}
F.~Borzumati and A.~Masiero,
Phys.\ Rev.\ Lett.\ {\bf 57} (1986) 961;
J.~Hisano, T.~Moroi, K.~Tobe, M.~Yamaguchi and T.~Yanagida,
Phys.\ Lett.\ B {\bf 357} (1995) 579;
J.~Hisano, T.~Moroi, K.~Tobe and M.~Yamaguchi,
Phys.\ Rev.\ D {\bf 53} (1996) 2442;
J.~Hisano and D.~Nomura,
Phys.\ Rev.\ D {\bf 59} (1999) 116005.

\bibitem{Farzan:2004qu}
Y.~Farzan and M.~E.~Peskin, hep-ph/0405214.

\bibitem{Ellis:2002fe}
J.~R.~Ellis, J.~Hisano, M.~Raidal and Y.~Shimizu,
Phys.\ Rev.\ D {\bf 66} (2002) 115013.

\bibitem{Okada}Y.~Okada, K.~Okumura and Y.~Shimizu,
Phys.\ Rev.\ D {\bf 58} (1998) 051901;
Phys.\ Rev.\ D {\bf 61} (2000) 094001.

\bibitem{deGouvea:2000cf}
A.~de Gouvea, S.~Lola and K.~Tobe,
Phys.\ Rev.\ D {\bf 63} (2001) 035004.


\bibitem{Ellis:2001xt}
J.~R.~Ellis, J.~Hisano, S.~Lola and M.~Raidal,
Nucl.\ Phys.\ B {\bf 621} (2002) 208.

\bibitem{Ellis:2001yz}
J.~R.~Ellis, J.~Hisano, M.~Raidal and Y.~Shimizu,
Phys.\ Lett.\ B {\bf 528} (2002) 86.

\bibitem{Hisano:2003bd}
For example, see J.~Hisano and Y.~Shimizu,
Phys.\ Lett.\ B {\bf 565} (2003) 183;
M.~Ciuchini, A.~Masiero, L.~Silvestrini, S.~K.~Vempati and O.~Vives,
Phys.\ Rev.\ Lett.\  {\bf 92} (2004) 071801.


\bibitem{Moroi:2000tk}
T.~Moroi, Phys.\ Lett.\ B {\bf 493} (2000) 366.

\bibitem{Dimopoulos:1994gj}
S.~Dimopoulos and L.~J.~Hall,
Phys.\ Lett.\ B {\bf 344} (1995) 185.

\bibitem{Barbieri:1995tw}
R.~Barbieri, L.~J.~Hall and A.~Strumia,
Nucl.\ Phys.\ B {\bf 445} (1995) 219;
R.~Barbieri, L.~J.~Hall and A.~Strumia,
Nucl.\ Phys.\ B {\bf 449} (1995) 437.


\bibitem{Romanino:1996cn}
A.~Romanino and A.~Strumia,
Nucl.\ Phys.\ B {\bf 490} (1997) 3.

\bibitem{Hisano:2004pw}
J.~Hisano, M.~Kakizaki, M.~Nagai and Y.~Shimizu,
hep-ph/0407169.

\bibitem{HS2}
J.~Hisano and Y.~Shimizu,
hep-ph/0406091.

\bibitem{Harris:jx}
P.~G.~Harris {\it et al.},
Phys.\ Rev.\ Lett.\  {\bf 82} (1999)  904.

\bibitem{Romalis:2000mg}
M.~V.~Romalis, W.~C.~Griffith and E.~N.~Fortson,
Phys.\ Rev.\ Lett.\  {\bf 86}  (2001) 2505.

\bibitem{Semertzidis:2003iq}
Y.~K.~Semertzidis {\it et al.}  [EDM Collaboration],
AIP Conf.\ Proc.\  {\bf 698} (2004)  200.

\bibitem{gw}
Y.~Grossman and M.~P.~Worah,
Phys.\ Lett.\ B {\bf 395} (1997) 241;
R.~Barbieri and A.~Strumia,
Nucl.\ Phys.\ B {\bf 508} (1997) 3.

\bibitem{HS}
J.~Hisano and Y.~Shimizu,
Phys.\ Lett.\ B {\bf 581} (2004) 224.

\bibitem{babar}
Talked by M.~Giorgi in ICHEP'04, August 16-22, 2004
Beijing, China, ({\sl http://ichep04.ihep.ac.cn/}).

\bibitem{belle}
Talked by Y.~Sakai in ICHEP'04, August 16-22, 2004
Beijing, China, ({\sl http://ichep04.ihep.ac.cn/}).


\end{thebibliography}
\end{document}